\documentclass[12pt]{iopart}
\usepackage{graphicx} 
\usepackage{dcolumn} 
\usepackage{amssymb}
\usepackage{iopams}
 \usepackage[colorlinks=true,citecolor=blue]{hyperref}

\def\sech#1{\mathrm{sech}#1}
\begin{document}

\title[QFT Cauchy]{Regular Quantum States on the Cauchy Horizon of a Charged Black Hole}
\author{Peter Taylor}
\address{School of Mathematical Sciences \& Centre for Astrophysics and Relativity, Dublin City University, Glasnevin, Dublin 9, Ireland}
\eads{\mailto{peter.taylor@dcu.ie}}
\date{\today}
\begin{abstract}
We consider the quantum stress-energy tensor of a massless scalar field near the Cauchy horizon interior to the Reissner-Nordstr\"om black hole spacetime. We construct the quantum state by considering the two-point function on a negative definite metric obtained by a double analytic continuation from the Lorentzian manifold, complexifying both the $t$ and polar coordinates. We enforce periodicity in the Euclideanized $t$ coordinate with periodicity equal to the reciprocal of the temperature of the Cauchy horizon, a necessary condition for avoiding a conical singularity at the inner horizon. We show by explicit construction that our quantum state satisfies the Hadamard condition on the Cauchy horizon. The expectation value of the quantum stress-energy tensor on the Cauchy horizon is given in closed form. \end{abstract}
\maketitle


\section{Introduction}
The Reissner-Nordstr\"om spacetime is a static spherically-symmetric solution of the Einstein-Maxwell field equations representing an electrically charged black hole. Unlike the Schwarzschild black hole, there are two coordinate singularities, one corresponding to the black hole event horizon and the other the Cauchy horizon inside the black hole. It is well known that this Cauchy horizon is unstable to classical perturbations \cite{McNamara1978a, Mcnamara1978b, Matzner1979, Gursel1979a, Gursel1979b, Gnedin1993, BradyPRL1995, BurkoPRL1997} where the inner horizon forms a null (weak) singularity, resulting in a metric which is continuous but not differentiable. This instability also holds more generally for perturbations inside rotating black holes \cite{OriPRL1992, OriPRD1999, OriPRL1999, BradyPRD1998}.

Less studied is the quantum back-reaction on the Cauchy horizon. Indeed, whether or not quantum effects significantly alter the classical instability remains an open problem. Assuming the Cauchy horizon is sufficiently far from the essential singularity, one can expect that the problem can be addressed in the semi-classical regime, treating the perturbations from quantum fields on the classical Reissner-Nordstr\"om spacetime. Even still this is a notoriously difficult problem, in part because the stress-energy tensor of the quantum fields, which act as the source in the semi-classical equations, is formally divergent and requires a suitable regularization prescription. While several authors have considered quantum fields near the inner horizon inside a black hole, as far as this author is aware, in all the prior body of work the field is taken to be in a quantum state that is singular on the Cauchy horizon, that is a state which does not satisfy the Hadamard condition there (see, for example, Ref.~\cite{WaldQFT} for a discussion of Hadamard states in quantum field theory in curved spacetimes). For example, in a 2D analogue of the Reissner-Nordstr\"om black hole, Birrell and Davies \cite{BirrellDaviesNature1978} showed that the regularized stress-energy tensor in the Hartle-Hawking state diverges on the Cauchy horizon. Similarly, in Ref.~\cite{Hiscock1980}, it is shown that the stress-energy tensor in the Unruh state diverges on the Cauchy horizon of a slowly rotating black hole, assuming the rotation parameter is continuous. More recently, a detailed analysis of the asymptotic behaviour of the divergence of the stress-energy tensor near the Cauchy horizon of the Reissner-Nordstr\"om black hole in both the Unruh and Hartle-Hawking states have been performed \cite{SelaPRD2018} as well as a numerical computation of the vacuum polarization on the interior \cite{LanirPRD2019}.

In this paper, we compute the regularized stress-energy tensor for a massless scalar field which is arbitrarily coupled to the background curvature near the Cauchy horizon of a Reissner-Nordstr\"om black hole. Importantly, we construct the field in a quantum state that is regular on this horizon, that is, a quantum state that satisfies the Hadamard condition on the inner horizon. It is widely accepted that only states that satisfy the Hadamard condition are physically meaningful so we believe that any meaningful conclusions drawn about the quantum back-reaction on the Cauchy horizon must be based on consideration only of Hadamard states. Moreover, the assumptions that underpin the semi-classical approximation are clearly violated near the Cauchy horizon for states which are singular there.

Our construction of a regular state involves employing Euclidean techniques, though in a novel way. In the usual approach to constructing the Hartle-Hawking state in a static black hole spacetime, it is convenient to perform a Wick rotation, which corresponds to complexifying the $t$ coordinate, and then imposing periodicity in the Euclidean time \cite{HartleHawkingPRD1976}. This results in a state which is thermal on the exterior, which has the same symmetries as the underlying spacetime and which is regular on the event horizon. On the black hole interior, the $t$ coordinate is spacelike and complexifying this coordinate results in a metric with a neutral signature. However, if in addition to complexifying the $t$ coordinate, we also complexify the polar coordinate $\theta\to \rmi \Theta$, then we retrieve a metric with a Euclidean signature (in fact, a negative definite metric but the overall sign is irrelevant). Like the standard Euclidean procedure, we impose periodicity in $\tau$, now the periodicity is related to the temperature of the Cauchy horizon rather than the event horizon. We show, by explicitly computing the regularized stress-energy tensor on the Cauchy horizon, that this state satisfies the Hadamard condition. In order to calculate the stress-energy tensor exactly on the horizon, a uniform asymptotic analysis of the radial modes is required. We develop a uniform asymptotic series for the radial modes which enables us to compute the stress-energy tensor in closed form.

The layout of the paper is as follows: In Sec.~\ref{sec:ext}, we briefly review the construction of the Euclidean two-point function for a scalar field in the Hartle-Hawking state on the exterior of the Reissner-Nordstr\"om spacetime. In Sec.~\ref{sec:int}, we construct the two-point function for a scalar field on the interior. The quantum state is defined by working on the negative definite spacetime obtained by a double analytic continuation. In Sec.~\ref{sec:VP}, we compute the vacuum polarization for the field in this quantum state and in Sec.~\ref{sec:RSET}, we compute the regularized stress-energy tensor in the state defined on this Euclidean section. The computation is completely analytical (modulo an arbitrary constant which encodes information about the quantum state and is calculated numerically) resulting in a closed form expression for the stress-energy tensor. This is made possible by a uniform asymptotic expansion for the radial modes which we discuss in detail in the Appendix.

\section{Review of the Hartle-Hawking State on the Exterior}
\label{sec:ext}
The Reissner-Nordstr\"om spacetime in spherical coordinates $(t, r, \theta, \phi)$ has line element
\begin{equation}
\fl
\varepsilon\,{\rmd}s^2 = -(1-2M/r+Q^{2}/r^{2}) {\rmd}t^2 +(1-2M/r+Q^{2}/r^{2})^{-1} {\rmd}r^2 +r^2 {\rmd} \theta^2 +  r^{2} \sin^{2}\theta {\rmd} \phi^{2}.
\end{equation}
This spacetime has two coordinate singularities at
\begin{equation}
	r_{\pm}=M\pm\sqrt{M^{2}-Q^{2}},
\end{equation}
and one essential curvature singularity at $r=0$. Assuming $M>Q$, the outer coordinate singularity $r_{+}$ represents a black hole event horizon while the inner coordinate singularity $r_{-}$ represents a Cauchy horizon. Before discussing how to construct the two-point function for a scalar field in a quantum state that is regular on the Cauchy horizon, let us first briefly review how to construct the two-point function in the Hartle-Hawking state \cite{HartleHawkingPRD1976} on the exterior $r\ge r_{+}$. This is not a pure state on the exterior, but a thermal state corresponding to the black hole in a thermal bath of radiation at the same temperature as the black hole. Since this is a thermal state, it is convenient to work with the Euclidean Green function, performing a Wick rotation of the temporal coordinate $t\rightarrow -i \tau$ and eliminating the conical singularity at $r=r_{+}$ by making $\tau$ periodic with period
$2\pi/\kappa_{+}$ where $\kappa_{+}=1/(2 r_{+})$ is the surface gravity of the black hole event horizon. Note that the signature of the metric is Euclidean only on the exterior. 


The field on the exterior satisfies the Klein Gordon equation with respect to the Euclideanized metric
\begin{equation}
\Box \varphi(\tau,r,\theta,\phi)=0,
\end{equation}
where $\Box$ denotes the wave operator with respect to the Euclidean metric. Note that the coupling of the field to the background curvature is irrelevant since the Ricci scalar vanishes on the Reissner-Nordstr\"om spacetime. The Klein Gordon equation can be solved by a separation of variables by writing
\begin{equation}
\varphi(\tau,r,\theta,\phi)\sim \rme^{i n \kappa_{+} \tau+i m\phi} P(\theta)\chi(r)
\end{equation}
where $P(\theta)$ is regular and satisfies 
\begin{equation}
\label{eq:legendre}
\Big\{\frac{1}{\sin\theta} \frac{{\rmd}}{{\rmd} \theta}\Big(\sin\theta\frac{{\rmd}}{{\rmd} \theta}\Big) -\frac{m^{2}}{\alpha^{2}\sin^{2}\theta}+\lambda(\lambda+1)\Big\}P(\theta)=0
\end{equation}
while $\chi(r)$ satisfies
\begin{equation}
\label{eq:radialhomogeneous}
\Big\{\frac{{\rmd}}{{\rmd} r}(r^2-2Mr+Q^{2})\frac{{\rmd}}{{\rmd}r} - \lambda(\lambda+1)-\frac{n^2 \kappa_{+}^2 r^4}{r^2-2Mr+Q^{2}}\Big\}\chi(r)= 0.
\end{equation}
The $\lambda(\lambda+1)$ term arises as the separation constant. The choice of $\lambda$ is arbitrary for $\varphi$ to satisfy the wave equation but is constrained by a choice of boundary conditions. In the exterior spacetime, one chooses regularity on the poles which implies $\lambda=l\in \mathbb{N}$, i.e the separation constant is $l(l+1)$. With this choice of $\lambda$, the angular functions are the standard associated Legendre function of integer degree and order,\textit{viz.},
\begin{equation}
P(\theta)=\mathsf{P}_{l}^{m}(\cos\theta),
\end{equation}
satisfying the normalization,
\begin{eqnarray}
\label{eq:norm}
\int_{-1}^{1}\mathsf{P}_{l}^{-m}(\cos\theta) \mathsf{P}_{l'}^{-m}(\cos\theta) d(\cos\theta) 
=\frac{2}{(2l+1)}\frac{\Gamma(l-m+1)}{\Gamma(l+m+1)}\delta_{ll'}. 
\end{eqnarray}

The periodicity of the Green function with respect to $(\tau-\tau')$ and $(\phi-\phi')$ with periodicity $2\pi/\kappa_{+}$ and $2\pi$, respectively, combined with Eq.(\ref{eq:norm}) imply the following mode-sum expression for the Green function
\begin{eqnarray}
\label{eq:greensfn}
G(x,x')=\frac{\kappa_{+}}{8\pi^{2}}\sum_{n=-\infty}^{\infty} {\rme}^{i n \kappa_{+} (\tau-\tau')}\sum_{l=0}^{\infty} (2l+1)\mathsf{P}_{l}(\cos\gamma)g_{n l}(r,r'),
\end{eqnarray}
where $\cos\gamma=\cos\theta\cos\theta'+\sin\theta\sin\theta'\cos\Delta\phi$ and $g_{n l}(r,r')$ satisfies the inhomogeneous equation,
\begin{equation}
\label{eq:radialr}
\fl
\Big\{\frac{{\rmd}}{{\rmd} r}(r^2-2Mr+Q^{2})\frac{{\rmd}}{{\rmd}r} - l(l+1)-\frac{n^2 \kappa_{+}^2 r^4}{r^2-2Mr+Q^{2}}\Big\}g_{nl}(r,r')=-\delta(r-r').\nonumber\\
\end{equation}

It is convenient to introduce a new dimensionless radial variable
\begin{equation}
\label{eq:eta}
\eta=\frac{r-M}{\alpha},\qquad \alpha=\sqrt{M^{2}-Q^{2}}.
\end{equation}
In terms of this new coordinate, the event horizon is located at $\eta=1$ while the Cauchy horizon is located at $\eta=-1$. The curvature singularity is at $\eta=-M/\alpha<-1$. Now the radial Green function in terms of $\eta$ assumes the form 
\begin{eqnarray}
\label{eq:radial}
\fl
\Big\{\frac{{\rmd}}{{\rmd}\eta}\Big((\eta^{2}-1)\frac{{\rmd}}{{\rmd}\eta}\Big)-l(l+1)-\frac{\alpha^{2}n^{2}\kappa_{+}^{2}(\eta+M/\alpha)^{4}}{(\eta^{2}-1)}\Big\}g_{n\lambda}(\eta,\eta')=-\frac{1}{\alpha }\delta(\eta-\eta').
\end{eqnarray}
For $n=0$, the two solutions of the homogeneous equation are the Legendre functions of the first and second kind, which we denote by $P_{l}(\eta)$ and $Q_{l}(\eta)$, respectively. For $n\ne 0$, the homogeneous equation cannot be solved in terms of known functions and must be solved numerically. We denote the two solutions that are regular on the horizon and infinity (or some outer boundary) by $p_{l}^{|n|}(\eta)$ and $q_{l}^{|n|}(\eta)$, respectively. A near-horizon Frobenius analysis for $n \neq 0$ shows that the indicial exponent is $\pm |n|/2$, and so we have the following asymptotic forms:
\begin{equation}
\label{eq:asymp}
\eqalign{
 p_{l}^{|n|}(\eta)\sim (\eta-1)^{|n|/2} \qquad\qquad &\eta\rightarrow 1, \cr
q_{l}^{|n|}(\eta) \sim (\eta-1)^{-|n|/2} &\eta \rightarrow 1.}
\end{equation}
Using these asymptotic forms in the Wronskian condition yields the appropriate normalization of the radial Green function:
 \begin{eqnarray}
 \label{eq:chi}
 g_{nl}(\eta,\eta') = 
\cases{
\displaystyle{\frac{1}{\alpha}} P_{l}(\eta_{<})Q_{l}(\eta_{>})&$n=0$, \\
\displaystyle \frac{1}{2|n|\alpha} p_{l}^{|n|}(\eta_{<}) q_{l}^{|n|}(\eta_{>})\qquad&$n\neq 0$.}
 \end{eqnarray}


\section{Green Function on the Interior}
\label{sec:int}
Turning now to the calculation of the two-point function on the interior of the black hole. In particular, we will consider the region between the Cauchy horizon and the event horizon. Like the Hartle-Hawking state, we will define the quantum state by employing Euclidean techniques. However, complexifying the $t$ coordinate results in a neutral signature metric on the interior between the Cauchy and event horizons. We can retrieve a metric of definite signature by further complexifying the polar coordinate by $\theta\to \rmi \Theta$. This results in a negative definite metric, though the overall sign is irrelevant. The quantum state on this spacetime is defined by constructing the two-point function for the scalar field on this double analytically continued metric and imposing regularity boundary conditions on the Cauchy horizon. We should note that this double analytic continuation was adopted by Candelas and Jensen \cite{CandelasPRD1986} to discuss the Feynman Green function on the interior of the Schwarzschild black hole. In practice, however, the authors constructed the two-point function on the interior by analytically continuing the exterior two-point function. This approach applied to the Reissner-Nordstr\"om back hole would result in a quantum state which is regular on the event horizon but singular on the Cauchy horizon.

Working again with the dimensionless variable $\eta$ defined by Eq. (\ref{eq:eta}), the analytically continued Reissner-Nordstr\"om metric is
\begin{equation}
\fl	\varepsilon\,ds^{2}=-\frac{(1-\eta^{2})}{(\eta+M/\alpha)^{2}}\rmd\tau^{2}-\alpha^{2}\frac{(\eta+M/\alpha)^{2}}{(1-\eta^{2})}\rmd\eta^{2}-\alpha^{2}(\eta+M/\alpha)^{2}(\rmd\Theta^{2}+\sinh^{2}\Theta\,\rmd\phi^{2}),
\end{equation}
where we are concerned with quantum effects on $-1\le\eta<1$ and specifically effects very close to the Cauchy horizon $\eta=-1$. In order to avoid a conical singularity at $\eta=-1$, we must enforce a periodicity on $\tau$, namely, $\tau=\tau+2\pi/\kappa_{-}$, where $\kappa_{-}$ is the surface gravity on the Cauchy surface. Assuming a separable basis,
\begin{equation}
	\varphi\sim e^{in\kappa_{-}\tau}e^{im \phi}P(\Theta)\chi(\eta),
\end{equation}
for solutions to the wave equation requires that $P(\Theta)$ satisfies
\begin{equation}
\label{eq:conical}
\Big\{\frac{1}{\sinh\Theta} \frac{{\rmd}}{{\rmd} \Theta}\Big(\sinh\Theta\frac{{\rmd}}{{\rmd} \Theta}\Big) -\frac{m^{2}}{\sinh^{2}\theta}-\nu(\nu+1)\Big\}P(\Theta)=0.
\end{equation}
The only choice of $\nu$ for which $P(\Theta)$ is square-integrable is $\nu=-1/2+\rmi \lambda$ for $\lambda$ real. For this choice, the mode functions are the conical (also referred to as Mehler or hyperboloidal) functions,
\begin{equation}
	P(\Theta)=P^{m}_{-1/2+\rmi \lambda}(\cosh\Theta).
\end{equation} 
These satisfy the orthogonality relation
\begin{equation}
\fl	\int_{1}^{\infty}P^{m}_{-1/2+\rmi \lambda}(z)	P^{m}_{-1/2+\rmi \lambda'}(z) \rmd z=\frac{(-1)^{m}\Gamma(\rmi\lambda+\frac{1}{2}+m)}{\lambda\,\tanh\pi\lambda\,\Gamma(\rmi\lambda+\frac{1}{2}-m)}\delta(\lambda-\lambda').
\end{equation}
Using these (appropriately normalized) basis modes to expand the Green function, and after employing a standard addition theorem for the conical functions \cite{gradriz}, we obtain,
\begin{equation}
	\label{eq:greensfnint}
	\fl G(x,x')=\frac{\kappa_{-}}{4\pi^{2}}\sum_{n=-\infty}^{\infty}\rme^{\rmi n \kappa_{-}\Delta\tau}\int_{0}^{\infty}\rmd\lambda\,\lambda\,\tanh\pi\lambda\,P_{-1/2+\rmi\lambda}(\cosh\Gamma)\mathsf{g}_{n\lambda}(\eta,\eta'),
\end{equation}
where $\cosh\Gamma=\cosh\Theta\cosh\Theta'-\sinh\Theta\sinh\Theta'\cos\Delta\phi$. The radial Green function $\mathsf{g}_{n\lambda}(\eta,\eta')$ satisfies
\begin{equation}
	\label{eq:radialeqn}
\fl\Bigg\{\frac{\rmd}{\rmd\eta}\Big((1-\eta^{2})\frac{\rmd}{\rmd\eta}\Big)-\lambda^{2}-\frac{1}{4}-\frac{\alpha^{2}n^{2}\kappa_{-}^{2}(\eta+M/\alpha)^{4}}{(1-\eta^{2})}\Bigg\}\mathsf{g}_{n\lambda}(\eta,\eta')=\frac{\delta(\eta-\eta')}{\alpha}.
\end{equation}
For $n=0$, the independent solutions to the homogeneous equation are $\mathsf{P}_{-1/2+\rmi\lambda}(\pm\eta)$ with
\begin{equation}
	\mathsf{P}_{-1/2+\rmi\lambda}(-\eta)=\frac{2}{\pi}\cosh \pi\lambda\,\Re\{\mathsf{Q}_{-1/2+\rmi \lambda}(\eta)\}
\end{equation}
being the solution regular on the Cauchy horizon since $\mathsf{P}_{-1/2+\rmi\lambda}(-\eta)\to 1$ as $\eta\to-1$. The appropriate normalization for these solutions is
\begin{equation}
	N= (1-\eta^{2})W\left\{\mathsf{P}_{-1/2+\rmi\lambda}(-\eta),\mathsf{P}_{-1/2+\rmi\lambda}(\eta)\right\}=-\frac{2}{\pi}\cosh \pi\lambda.
\end{equation}
For $n\ne 0$, the solutions cannot be given in terms of known functions but must be solved numerically. We will denote the solution regular on the Cauchy horizon by $\mathsf{q}^{|n|}_{\lambda}(\eta)$ and the solution which diverges there by $\mathsf{p}^{|n|}_{\lambda}(\eta)$. Analysis of the Frobenius series about $\eta=-1$ yields the following asymptotic behaviour for these solutions
\begin{equation}
\label{eq:asympinterior}
\eqalign{
 \mathsf{q}_{l}^{|n|}(\eta)\sim (1+\eta)^{|n|/2} \qquad\qquad &\eta\rightarrow -1, \cr
\mathsf{p}_{l}^{|n|}(\eta) \sim (1+\eta)^{-|n|/2} &\eta \rightarrow -1.}
\end{equation}
The solution to the inhomogeneous equation is the normalized product,
\begin{equation}
	\mathsf{g}_{n\lambda}(\eta,\eta')=\frac{\mathsf{q}^{|n|}_{\lambda}(\eta_{<})\mathsf{p}^{|n|}_{\lambda}(\eta_{>})}{\alpha\,N_{n}}
\end{equation}
where $N_{n}=(1-\eta^{2})W\{\mathsf{q},\mathsf{p}\}$. Using the asymptotic forms above to compute the Wronskian gives $N_{n}=-2|n|$. Finally, we can write the solution for all $n$ as
 \begin{eqnarray}
 \label{eq:gradialint}
 \mathsf{g}_{n\lambda}(\eta,\eta') = 
\cases{
\displaystyle{-\frac{\pi}{2\alpha\,\cosh \pi\lambda}} \mathsf{P}_{-1/2+\rmi\lambda}(-\eta_{<})\,\mathsf{P}_{-1/2+\rmi\lambda}(\eta_{>})&$n=0$, \\
\displaystyle -\frac{1}{2|n|\alpha} \mathsf{q}_{\lambda}^{|n|}(\eta_{<}) \mathsf{p}_{l}^{|n|}(\eta_{>})\qquad&$n\neq 0$.}
 \end{eqnarray}
 
\section{Vacuum Polarization on the Cauchy Horizon}
\label{sec:VP}
In this section, we compute the vacuum polarization for a massless scalar field on the Cauchy horizon of the Reissner-Nordstr\"om black hole. The field is assumed to be in the quantum state defined by the double analytic continuation procedure described in the previous section. The vacuum polarization for the field in this state is defined to be the coincidence limit of the regularized two-point function,
\begin{equation}
	\langle \hat{\varphi}^{2}\rangle=\lim_{x'\to x}\left[G(x,x')-G_{\mathrm{S}}(x,x')\right]
\end{equation}
where $G_{\mathrm{S}}(x,x')$ is a parametrix for the wave operator, symmetric in $x$ and $x'$ and is constructed only from the geometry through the metric and its derivatives (see, for example, Ref.~\cite{WaldQFT}). We take $G_{\mathrm{S}}(x,x')$ to be a Hadamard parametrix which we define below. Since we are interested in computing the vacuum polarization exactly on the Cauchy horizon, consideration of the asymptotic forms (\ref{eq:asympinterior}) implies that taking one point on the horizon means that all the $n\neq 0$ modes vanish. Therefore the Green function with one point on the Cauchy horizon is independent of $\tau$ and reduces to
\begin{eqnarray}
\label{eq:greensfnnzero}
\fl
G(\eta,\Omega;-1,\Omega')=-\frac{\kappa_{-}}{8 \pi^{2} \alpha} \int_{\lambda=0}^{\infty} \lambda\,\frac{\pi\,\tanh \pi\lambda}{\cosh \pi\lambda} P_{-1/2+\rmi \lambda}(\cosh\Gamma)\,\mathsf{P}_{-1/2+\rmi\lambda}(\eta),
\end{eqnarray}
where $\Omega=(\Theta, \phi)$. This integral is essentially an analytic continuation of the Heine identity and can be performed in closed form \cite{CandelasPRD1986} yielding,
\begin{equation}
	\label{eq:Heine}
	\int_{\lambda=0}^{\infty} \lambda\,\frac{\pi\,\tanh \pi\lambda}{\cosh \pi\lambda} P_{-1/2+\rmi \lambda}(z)\,\mathsf{P}_{-1/2+\rmi\lambda}(y)=\frac{1}{z+y}.
\end{equation}
Applying this result gives
\begin{eqnarray}
\label{eq:Gclosed}
G(\eta,\Omega;-1,\Omega')=-\frac{\kappa_{-}}{8 \pi^{2} \alpha} \frac{1}{(\cosh\gamma+\eta)}.
\end{eqnarray}
This, of course, diverges in the coincidence limit $\eta\to -1$ and $\Omega'\to\Omega$, as expected. 

To regularize, we adopt the Hadamard regularization prescription. This relies on the universal Hadamard singularity structure of the two-point function for $x$ and $x'$ sufficiently close together that they are connected by a unique geodesic. The singularities are encoded in the so-called Hadamard parametrix
\begin{equation}
	\label{eq:Gsing}
	G_{\mathrm{S}}(x,x')=\frac{1}{4\pi^{2}}\left(\frac{\Delta^{1/2}(x,x')}{2\sigma(x,x')}+V(x,x')\log(2\sigma(x,x')/\ell^{2})\right),
\end{equation}
where $2\sigma(x,x')$ is the squared geodesic distance between $x$ and $x'$ with respect to the Euclideanized spacetime, $\Delta(x,x')$ is the Van Vleck-Morrette determinant, $V(x,x')$ is a regular, symmetric biscalar which satisfies the same Klein-Gordon equation satisfied by our scalar field, and $\ell$ is an arbitrary lengthscale required to make the log term dimensionless. Not all quantum states have a corresponding two-point function with this universal Hadamard structure, though only those that satisfy this Hadamard condition are generally considered physically meaningful \cite{WaldQFT}. A key problem addressed in this paper is how to construct a quantum state that satisfies the Hadamard condition near the Cauchy horizon.

Each of the biscalars in (\ref{eq:Gsing}) can be covariantly Taylor expanded about one of the points (for high-order covariant expansions in arbitrary dimensions, see for example Ref.~\cite{DecaniniFolacciHadamardRen}). Treating the separation between the points as formally $\mathcal{O}(\epsilon)=\mathcal{O}(\Delta x)=\mathcal{O}(\sigma^{;a})$, for the Van Vleck-Morrette determinant, we get
\begin{eqnarray}
	\Delta^{1/2}=1+\frac{1}{12}R_{a'b'}\sigma^{;a'}\sigma^{;b'}+\mathcal{O}(\epsilon^{3}).
\end{eqnarray}
Considering massless fields on a background geometry with vanishing scalar curvature implies $V=\textrm{O}(\epsilon^{2})$ and hence the tail term does not contribute in the coincidence limit. There are also standard coordinate expansions for these biscalars but they turn out to be useless in this context since the metric in the coordinates we have adopted is singular on the Cauchy horizon. Though coordinates exist in which the metric is regular on this horizon, the transformation cannot be given explicitly. Regardless, since we are ultimately interested in computing these biscalars in the coincidence limits, we can simplify things considerably by separating only in the radial direction whence the terms in the covariant expansion can be computed exactly. For the world function $\sigma(x,x')$, we can compute this for radial separation by directly integrating the line-element with $\rmd\tau=\rmd\Theta=\rmd\phi=0$. Assuming $-1<\eta'<\eta<1$, this gives
\begin{equation}
	\sqrt{-2\sigma}=s=\int_{\eta'}^{\eta}\frac{(\alpha\eta''+M)}{\sqrt{1-\eta''^{2}}}d\eta''.
\end{equation}
Moreover, since we are only interested in the separation along radial directions, we have $\sigma^{;a'}\equiv 0$ except 
\begin{equation}
\sigma^{;\eta'}=\frac{\sqrt{1-\eta'^{2}}}{(\alpha\eta'+M)}\,s.
\end{equation}
Putting this together gives
\begin{eqnarray}
	\label{eq:Gsradial} G_{\mathrm{S}}(\eta,\eta')&=&\frac{\Delta^{1/2}}{8\pi^{2}\sigma}+\mathcal{O}(\epsilon^{2}\ln\epsilon)=-\frac{1}{4\pi^{2}s^{2}}-\frac{1}{48\pi^{2}}R_{\eta'\eta'}\frac{(\sigma^{;\eta'})^{2}}{s^{2}}+\mathcal{O}(\epsilon)\nonumber\\
	&=&-\frac{1}{4\pi^{2}s^{2}}-\frac{Q^{2}}{48\pi^{2}(\alpha\eta'+M)^{4}}+\mathcal{O}(\epsilon).
\end{eqnarray}
Now taking one point on the horizon, $\eta'=-1$, and taking $\eta=-1+\epsilon$ for some $\epsilon>0$ gives
\begin{equation}
	s=\int_{-1}^{-1+\epsilon}\frac{(\alpha\eta+M)}{\sqrt{1-\eta^{2}}}d\eta=-\alpha\sqrt{\epsilon(2-\epsilon)}+M\,\arccos(1-\epsilon).
\end{equation}
Substituting this into (\ref{eq:Gsradial}) and expanding in $\epsilon$ gives, after some algebra,
\begin{equation}
	\label{eq:Gsplit}
	G_{\mathrm{S}}=-\frac{1}{8\pi^{2}r_{-}^{2}\epsilon}+\frac{\kappa_{-}}{24\pi^{2}r_{-}}+\mathcal{O}(\epsilon),
\end{equation}
where we note that $\kappa_{-}=-f'(r_{-})/2=\alpha/r_{-}^{2}$. Taking $\Omega'\to\Omega$ and $\eta=-1+\epsilon$ in Eq.~(\ref{eq:Gclosed}) gives simply
\begin{equation}
	G=-\frac{\kappa_{-}}{8\pi^{2}\alpha\epsilon}=-\frac{1}{8\pi^{2}r_{-}^{2}\epsilon}.
\end{equation}
Finally, subtracting the Hadamard parametrix (\ref{eq:Gsplit}) from this and taking the coincidence limit $\epsilon\to 0$ gives the vacuum polarization
\begin{equation}
	\langle \hat{\varphi}^{2}\rangle=-\frac{\kappa_{-}}{24\pi^{2}r_{-}}.
\end{equation}
It is worth noting that the sign of the vacuum polarization on the Cauchy horizon is negative, in contrast to the vacuum polarization on the event horizon for a scalar field in the Hartle-Hawking state. Furthermore, we note that the vacuum polarization is regular on the Cauchy horizon in this quantum state, and hence this state satisfies the Hadamard condition. We still need to show that the same is true for the stress-energy tensor in this state, a significantly more involved calculation. This is computed in the following section.

\section{The Regularized Stress-Energy Tensor on the Cauchy Horizon}
\label{sec:RSET}
In this section, we calculate the regularized expectation value of the stress-energy tensor for a massless, arbitrarily coupled scalar field on the Cauchy horizon. This is the quantity of physical interest in obtaining the back-reaction on the spacetime geometry via the semi-classical field equations. 
%

For a massless scalar field propagating in a Ricci-flat background, we have the following expression for the classical stress-energy tensor
\begin{eqnarray}
\label{eq:Tclassical}
\fl
T^{a}{}_{b} =(1-2\xi) g^{a c}\varphi_{;c}\varphi_{;b}+(2\xi-\case{1}{2})\delta^{a}{}_{b}g^{c d}\varphi_{;c}\varphi_{;d}-2\xi g^{a c}\varphi \varphi_{;c b}+2\xi\delta^{a}{}_{b}\varphi\Box\varphi+\xi\,R^{a}{}_{b}\varphi^{2}.
\end{eqnarray}
In the point-splitting approach to regularization \cite{ChristensenPointSplit}, of which the Hadamard prescription is a variant, we write this tensor as a coincidence limit of a bi-tensor,
\begin{equation}
T^{a}{}_{b} = [\hat{D}^{a}{}_{b}(\varphi(x)\varphi(x'))]\equiv \lim_{x'\rightarrow x} \hat{D}^{a}{}_{b}(\varphi(x)\varphi(x'))
\end{equation}
where $\hat{D}^{a}{}_{b}=\hat{D}^{a}{}_{ b}(x,x')$ is a differential operator which may be defined in any way provided it gives (\ref{eq:Tclassical}) in the coincidence limit. We shall adopt the following definition,
\begin{equation}
\label{eq:stresstensorop}
\fl
 \hat{D}^{a}{}_{ b} = (1-2\xi) g^{a c'}\nabla_{b}\nabla_{c'}+(2\xi-\case{1}{2})\delta^{a}{}_{b}g^{c d'}\nabla_{c}\nabla_{d'}-2\xi g^{a c}\nabla_{c}\nabla_{b}+2\xi\delta^{a}{}_{b}\nabla_{c}\nabla^{c}+\xi\,R^{a}{}_{b}
\end{equation}
where $g^{a b'}$ are the bivectors of parallel transport. A well-known problem with adopting this definition is that the renormalized quantum stress-energy tensor is no longer conserved, though this is easily remedied by adding an appropriate factor of $v_{1}(x)=\left[V_{1}(x,x')\right]$. This corresponds to a redefinition of the arbitrary lengthscale $\ell$ in the Hadamard parametrix (\ref{eq:Gsing}). Taking this into consideration, we define the quantum expectation value of the stress-energy tensor for the field in our quantum state to be \cite{BrownOttewill1986}
\begin{equation}
	\label{eq:TrenDef}
	\langle \hat{T}^{a}{}_{b}\rangle=\lim_{x'\to x}\hat{D}^{a}{}_{b}\left(G(x,x')-G_{\mathrm{S}}(x,x')\right)+\frac{1}{4\pi^{2}}v_{1}(x)\delta^{a}{}_{b},
\end{equation}
where $G(x,x')$ is the Green function given by (\ref{eq:greensfnint}) and (\ref{eq:gradialint}) while $G_{\mathrm{S}}(x,x')$ is the Hadamard parametrix given by (\ref{eq:Gsing}). 

We focus first on the $\hat{D}^{a}{}_{b}G(x,x')$ term. Since we are concerned with calculating the stress-energy tensor exactly on the Cauchy horizon, it is most convenient to split in the radial direction as we did for the vacuum polarization above. In what follows, we assume, without loss of generality that $x'<x$ and then we consider taking the inner point $x'$ to lie on the horizon. Things are more complicated than in the calculation of the vacuum polarization however since we must now consider derivatives of the Green function, and taking $x'$ to the horizon or taking partial coincidence limits must be postponed until the derivatives have been performed. We will also need expansions of the bivectors of parallel transport. Fortunately, for radial separation, these are trivially obtained in closed form; in $(\tau,\eta,\Theta,\phi)$ coordinates, we have:
\begin{eqnarray}
\fl
g_{\tau\tau'}=-\frac{\sqrt{1-\eta^{2}}\sqrt{1-\eta'^{2}}}{(\eta+M/\alpha)(\eta'+M/\alpha)} &\qquad
g_{\eta \eta'}=-\alpha^{2}\frac{(\eta+M/\alpha)(\eta'+M/\alpha)}{\sqrt{1-\eta^{2}}\sqrt{1-\eta'^{2}}} \nonumber\\
\fl g_{\Theta\Theta'}= -\alpha^{2}(\eta+M/\alpha)(\eta'+M/\alpha) &\qquad
g_{\phi\phi'} = -\alpha^{2}(\eta+M/\alpha)(\eta'+M/\alpha) \sinh^{2}\Theta.
\end{eqnarray}

%
Examination of (\ref{eq:stresstensorop}) reveals that there is essentially two types of terms we need to evaluate in order to compute $\hat{D}^{a}{}_{b}G(x,x')$: those of the form $g^{a c'}G_{;c' b}$ and those of the form $g^{a c}G_{;c b}$.

Considering the latter case first. For such terms, we have two derivatives at the same spacetime point, which will involve the Christoffel symbols,
\begin{equation}
\label{eq:gcovariant}
G_{;a b} = G_{, a b}-\Gamma^{c}_{a b} G_{,c},
\end{equation}
where the Christoffel symbols in these coordinates are
\begin{eqnarray}
\label{eq:christoffel}
\fl \Gamma^{\tau}_{\tau\eta}=-\frac{\alpha+M\,\eta}{(M+\alpha\,\eta)(1-\eta^{2})} &\qquad
\Gamma^{\eta}_{\tau\tau}=\alpha^{2}\frac{(\alpha+M\,\eta)(1-\eta^{2})}{(M+\alpha\,\eta)^{5}}\nonumber\\
\fl\Gamma^{\eta}_{\eta\eta}=\frac{\alpha+M\,\eta}{(M+\alpha\,\eta)(1-\eta^{2})} &\qquad \Gamma^{\eta}_{\phi\phi}=\sinh^{2}\Theta\,\Gamma^{\eta}_{\Theta\Theta}=-\alpha\,\sinh^{2}\Theta\frac{(1-\eta^{2})}{M+\alpha\,\eta)}\nonumber\\
\fl\Gamma^{\Theta}_{\eta\Theta}=\Gamma^{\phi}_{\eta\phi}=\frac{\alpha}{M+\alpha\,\eta} &\qquad \Gamma^{\Theta}_{\phi\phi}=-\sinh^{2}\Theta\,\Gamma^{\phi}_{\Theta\phi}=-\sinh\Theta\,\cosh\Theta,
\end{eqnarray}
with all other coefficients being zero. Notwithstanding the extra term involving the Christoffel symbols, things are significantly easier when the two derivatives are taken at the same spacetime point since only the zero frequency mode contributes in the limit where one point is taken to the Cauchy horizon. To see this, note that we can always choose the derivative to act on the outer point, which we have chosen to be $x$ without loss of generality, and it is clear that this will not affect the asymptotics (\ref{eq:asympinterior}) at the inner spacetime point $x'$. In particular, taking $\eta'\rightarrow -1$ and using the asymptotic forms (\ref{eq:asympinterior}) immediately implies that all the modes vanish except the $n=0$ term. Since we have a closed form representation of the $n=0$ mode, each of the terms of the form $[g^{ac}G_{;c b}]$ can be obtained by directly differentiating (\ref{eq:Gclosed}). Performing the derivatives and using the appropriate Christoffel symbols, followed by taking the partial coincidence limits and expanding about the Cauchy horizon, we obtain
  \begin{eqnarray}
  \label{eq:Gtautau}
 \fl
\ \ [g^{\tau\tau}G_{;\tau\tau}]_{\mathrm{r_{-}}}=\frac{1}{8\pi^{2}r_{-}^{4}}\left\{-\frac{1}{(\eta+1)^{2}}+\frac{(M+3\alpha)}{r_{-}(\eta+1)}-\frac{3\alpha(2\alpha+M)}{r_{-}^{2}}\right\} +\Or(\eta+1),
 \end{eqnarray}
 \begin{eqnarray}
 \label{eq:Grr}
 \fl
\ \ [g^{\eta \eta}G_{;\eta\eta}]_{\mathrm{r_{-}}}=\frac{1}{8\pi^{2}r_{-}^{4}}\left\{\frac{3}{(\eta+1)^{2}}-\frac{(M+3\alpha)}{r_{-}(\eta+1)}+\frac{\alpha(2\alpha+M)}{r_{-}^{2}}\right\} +\Or(\eta+1),
 \end{eqnarray}
  \begin{eqnarray}
  \label{eq:Gthetatheta}
 \fl
\ \ [g^{\Theta\Theta}G_{;\Theta\Theta}]_{\mathrm{r_{-}}}=[g^{\phi\phi}G_{;\phi\phi}]_{\mathrm{r_{-}}}=\frac{1}{8\pi^{2}r_{-}^{4}}\left\{-\frac{1}{(\eta+1)^{2}}+\frac{\alpha(M+2\alpha)}{r_{-}^{2}}\right\} +\Or(\eta+1),
 \end{eqnarray}
where we have adopted square bracket notation $[..]_{r_{-}}$ to indicate that we have taken the partial coincidence limit $(\tau'\rightarrow \tau,\eta'\rightarrow -1,\Theta' \rightarrow \Theta,\phi'\rightarrow \phi)$. As a simple check of these expansions, one can see that adding these gives $\Box G=0$ up to the order of our expansions.

Turning now to terms of the form $g^{a c'}G_{;bc'}$. Such terms involve a covariant derivative at each spacetime point but since $G(x,x')$ is a scalar at both $x$ and $x'$, we are in fact only dealing with partial derivatives. For the angular terms $g^{\Theta\Theta'}G_{;\Theta\Theta'}$ and $g^{\phi\phi'}G_{;\phi\phi'}$, it is clear from the asymptotic forms (\ref{eq:asympinterior}) that taking $x'$ to lie on the horizon means that only the $n=0$ terms will contribute in the limit $\eta'\rightarrow -1$. Therefore, we can differentiate directly Eq.(\ref{eq:Gclosed}), take partial coincidence limits and expand about the horizon to obtain
\begin{eqnarray}
\label{eq:Gthetathetap}
\fl
[g^{\Theta\Theta'} G_{;\Theta\Theta'}]_{r_{-}}=[g^{\phi\phi'} G_{;\phi\phi'}]_{r_{-}}=\frac{1}{8\pi^{2}r_{-}^{4}}\left\{\frac{1}{(\eta+1)^{2}}-\frac{\alpha}{r_{-} (\eta+1)}+\frac{\alpha^{2}}{r_{-}^{2}}\right\}+\Or(\eta+1).
\end{eqnarray}

For $g^{\tau\tau'}G_{;\tau\tau'}$ and $g^{\eta \eta'}G_{;\eta \eta'}$, we must differentiate the full Green's function given by Eqs.(\ref{eq:greensfnint}) and (\ref{eq:gradialint}) before we can take $x'$ to lie on the Cauchy horizon. Considering $g^{\tau\tau'}G_{;\tau\tau'}$ first, differentiating and splitting only in the radial direction gives
\begin{eqnarray}
\fl
[g^{\tau\tau'}G_{;\tau\tau'}]_{r}= -\frac{(\eta+M/\alpha)(\eta'+M/\alpha)}{\sqrt{1-\eta^{2}}\sqrt{1-\eta'^{2}}}\frac{\kappa_{-}}{4\pi^{2}}\sum_{n=-\infty}^{\infty} n^{2}\kappa_{-}^{2}
\int_{0}^{\infty} \rmd\lambda\,\lambda\,\tanh\pi\lambda\,\mathsf{g}_{n\lambda}(\eta,\eta').
 \end{eqnarray}
 Trivially, the $n=0$ term will vanish. Moreover, using the asymptotic forms Eq.(\ref{eq:asympinterior}), we see that
 \begin{equation}	 \fl[g^{\tau\tau'}\mathsf{g}_{n\lambda}(\eta,\eta')]_{r}\sim\frac{r_{-}}{2\sqrt{2}\,\alpha^{2}|n|}\frac{(\eta+M/\alpha)\mathsf{p}^{|n|}_{\lambda}(\eta)}{\sqrt{1-\eta^{2}}}(1+\eta')^{|n|/2-1/2},\qquad \eta'\to-1,
\end{equation}
so that only the $n=\pm 1$ modes will be non-zero in the limit $\eta'\to-1$. Hence, taking this limit yields,
 \begin{eqnarray}
 \label{eq:Gtautaup}
 [g^{\tau\tau'} G_{;\tau\tau'}]_{r_{-}} =\frac{\kappa_{-}^{3}r_{-}}{8\sqrt{2} \pi^{2}\alpha^{2}} \frac{(\eta+M/\alpha)}{\sqrt{1-\eta^{2}}} F(\eta),
 \end{eqnarray}
 where
 \begin{equation}
 \label{eq:FDef}
 F(\eta)\equiv\int_{0}^{\infty}2 \lambda\,\tanh\pi\lambda\,\mathsf{p}^{1}_{\lambda}(\eta)\,\rmd\lambda.
 \end{equation}
We require a series expansion of $F(\eta)$ about the Cauchy horizon $\eta=-1$. To achieve this, however, involves a uniform asymptotic analysis of the radial solution $\mathsf{p}^{1}_{\lambda}(\eta)$, uniformly valid for both $\eta\sim-1$ and arbitrarily large $\lambda$. The precise details of this analysis are somewhat technical and are deferred to the appendix where it is shown that,
\begin{eqnarray}
	\label{eq:FSeries}
	 F(\eta)&=&\frac{2}{(1+\eta)^{3/2}}-\frac{(M+3\alpha)}{2 r_{-}(1+\eta)^{1/2}}-\left(\frac{(M+3\alpha)^{2}}{16r_{-}^{2}}-B\right)(1+\eta)^{1/2}\nonumber\\
	&+&\Or\left((\eta+1)^{3/2}\ln(1+\eta)\right),
\end{eqnarray}
where $B$ is, as of yet, an unspecified constant. To interpret this constant, note that $F(\eta)$ requires some input about the quantum state, or equivalently, about the boundary conditions imposed on the radial modes. In our asymptotic series which we outline in the appendix, information about the choice of boundary conditions is encoded in this constant $B$. For now, we make no particular choice. Substituting (\ref{eq:FSeries}) into (\ref{eq:Gtautaup}) gives
\begin{equation}
 [g^{\tau\tau'} G_{;\tau\tau'}]_{r_{-}}=\frac{1}{8\pi^{2}r_{-}^{4}}\left\{\frac{1}{(\eta+1)^{2}}+\left(\frac{B}{2}-\frac{\alpha(M+2\alpha)}{2 r_{-}^{2}}\right)\right\}+\Or(\eta+1).
\end{equation}

A similar argument to the one above can be employed to obtain an expansion for $[g^{\eta \eta'}G_{;\eta\eta'}]_{r_{-}}$. We have
 \begin{eqnarray}
\fl [g^{\eta \eta'}G_{;\eta\eta'}]_{r}= \frac{\sqrt{1-\eta^{2}}\sqrt{1-\eta'^{2}}}{\alpha^{2}(\eta+M/\alpha)(\eta'+M/\alpha)}\frac{\kappa_{-}}{4\pi^{2}}\sum_{n=-\infty}^{\infty}\int_{0}^{\infty} \rmd\lambda\,\lambda\,\tanh\pi\lambda\,\frac{\partial^{2}}{\partial \eta \partial \eta'}\mathsf{g}_{n\lambda}(\eta,\eta'). 
 \end{eqnarray}
Now the $n=0$ mode vanishes when we take $\eta'$ on the Cauchy horizon by merit of the fact that
\begin{equation}
\sqrt{1-\eta'^{2}}\frac{\rmd \mathsf{P}_{-1/2+\rmi\lambda}(-\eta')}{\rmd \eta'}=\mathsf{P}^{1}_{-1/2+\rmi\lambda}(-\eta') \rightarrow 0 \qquad \mathrm{as} \qquad \eta'\rightarrow -1.
\end{equation}
Moreover, using the asymptotic forms Eq.(\ref{eq:asympinterior}), we have for $n\ne 0$,
\begin{equation}
\fl\left[g^{\eta\eta'}\frac{\partial}{\partial\eta\partial\eta'}\mathsf{g}_{n\lambda}(\eta,\eta')\right]_{r}\sim \frac{|n|}{2\sqrt{2}\alpha^{2}r_{-}}\frac{\sqrt{1-\eta^{2}}}{(\eta+M/\alpha)}\frac{\partial\mathsf{p}^{|n|}_{\lambda}(\eta)}{\partial\eta}(\eta'+1)^{|n|/2-1/2},\qquad \eta'\to-1,
\end{equation}
which implies that all but the $n=\pm1$ terms vanish in the limit where one point is taken to the Cauchy horizon. Taking this limit yields
\begin{eqnarray}
\label{eq:Grrp}
[g^{\eta\eta'}G_{;\eta\eta'}]_{\mathrm{r_{-}}}=\frac{\kappa_{-}}{8\sqrt{2} \pi^{2}\alpha^{2} r_{-}}\frac{\sqrt{1-\eta^{2}}}{(\eta+M/\alpha)}\frac{\rmd F(\eta)}{\rmd \eta},
 \end{eqnarray}
 where the expansion for $F(\eta)$ about the Cauchy horizon is given by Eq.~(\ref{eq:FSeries}). Putting these together gives
\begin{eqnarray}
\fl	\left[g^{\eta \eta'}G_{;\eta \eta'}\right]_{r_{-}}=\frac{1}{8\pi^{2} r_{-}^{4}}\left\{-\frac{3}{(1+\eta)^{2}}+\frac{M+3\alpha}{r_{-}(1+\eta)}+\left(\frac{B}{2}-\frac{3\alpha(M+2\alpha)}{2 r_{-}^{2}}\right)\right\}+\Or(1+\eta).\nonumber\\
\end{eqnarray}

Now, defining $\langle \hat{T}^{a}{}_{b}\rangle_{\mathrm{unreg}}\equiv \left[\hat{D}^{a}{}_{b}G\right]_{r_{-}}$, with $\hat{D}^{a}{}_{b}$ defined by Eq.~(\ref{eq:stresstensorop}), we can compute
\begin{eqnarray}
	\label{eq:tunren}
\fl	\langle \hat{T}^{\tau}{}_{\tau}\rangle_{\mathrm{unreg}}&=&\frac{1}{8\pi^{2}r_{-}^{4}}\Bigg\{\frac{1}{(\eta+1)^{2}}+\frac{M(2\xi-1)-\alpha(6\xi+1)}{2\,r_{-}(\eta+1)}\nonumber\\
\fl&&+\left(B\,\xi+\frac{\alpha M(1+6\xi)+20\alpha^{2}\xi}{2 r_{-}^{2}}\right)\Bigg\}+\Or(1+\eta)\nonumber\\
\fl	\langle \hat{T}^{\eta}{}_{\eta}\rangle_{\mathrm{unreg}}&=&\frac{1}{8\pi^{2}r_{-}^{4}}\Bigg\{-\frac{3}{(\eta+1)^{2}}+\frac{M(6\xi+1)+\alpha(6\xi+5)}{2\,r_{-}(\eta+1)}\nonumber\\
\fl&&+\left(B\,\xi-\frac{\alpha\,M(1+6\xi)+4\alpha^{2}(1+\xi)}{2 r_{-}^{2}}\right)\Bigg\}+\Or(1+\eta)\nonumber\\
\fl	\langle \hat{T}^{\Theta}{}_{\Theta}\rangle_{\mathrm{unreg}}&=&\langle\hat{T}^{\phi}{}_{\phi}\rangle_{\mathrm{unreg}}=\frac{1}{8\pi^{2}r_{-}^{4}}\Bigg\{\frac{1}{(\eta+1)^{2}}+\frac{(M+3\alpha)(2\xi-1)}{2\,r_{-}(\eta+1)}\nonumber\\
\fl&&+\left(\frac{B}{2}(4\xi-1)+\frac{\alpha\,M(1-6\xi)+2\alpha^{2}(1-5\xi)}{r_{-}^{2}}\right)\Bigg\}+\Or(1+\eta).
\end{eqnarray}

The geometrical subtraction terms are found by obtaining a series expansion for the differential operator (\ref{eq:stresstensorop}) acting on the Hadamard parametrix and taking appropriate partial coincidence limits with one point placed on the Cauchy horizon. These are independent of the quantum state under consideration. Defining $\langle\hat{T}^{a}{}_{b}\rangle_{\mathrm{S}}\equiv \left[\hat{D}^{a}{}_{b}G_{\mathrm{S}}(x,x')\right]_{r_{-}}$, the results are
\begin{eqnarray}
\label{eq:tdiv}
\fl 
\langle \hat{T}^{\tau}{}_{ \tau} \rangle_{\mathrm{S}} &=& \frac{1}{8\pi^{2}r_{-}^{4}}\Bigg\{\frac{1}{(\eta+1)^{2}}+\frac{M(2\xi-1)-\alpha(6\xi+1)}{2 r_{-}(\eta+1)}\nonumber\\
\fl&&+\frac{1}{360 r_{-}^{2}}\left(51 M^{2}-2 M \alpha (180\xi-673)+\alpha^{2}(1080\xi+2171\right)\Bigg\}+\Or(\eta+1),\nonumber\\
\fl\langle\hat{T}^{\eta}{}_{\eta}\rangle_{\mathrm{S}}&=&\frac{1}{8\pi^{2}r_{-}^{4}}\Bigg\{-\frac{3}{(\eta+1)^{2}}+\frac{M(6\xi+1)+\alpha(6\xi+5)}{2\,r_{-}(\eta+1)}\nonumber\\
\fl&&-\frac{1}{360 r_{-}^{2}}\left(141 M^{2}+2 M \alpha (540 \xi+1811)+ \alpha^{2}(360\xi+7189)\right)\Bigg\}+\Or(1+\eta)\nonumber\\
\fl	\langle \hat{T}^{\Theta}{}_{\Theta}\rangle_{\mathrm{S}}&=&\langle\hat{T}^{\phi}{}_{\phi}\rangle_{\mathrm{S}}=\frac{1}{8\pi^{2}r_{-}^{4}}\Bigg\{\frac{1}{(\eta+1)^{2}}+\frac{(M+3\alpha)(2\xi-1)}{2\,r_{-}(\eta+1)}\nonumber\\
\fl&&+\frac{1}{360 r_{-}^{2}}\left(51 M^{2}-10 M \alpha (108 \xi-145)- \alpha^{2}(1080\xi-2707)\right)\Bigg\}+\Or(1+\eta).\nonumber\\
\fl
\end{eqnarray}
We note the absence of logarithmic singularities in these expressions. The coefficient of the logarithmic term in the Hadamard parametrix is the biscalar $V(x,x')$ which for a massless scalar field in Reissner-N{\"o}rdstrom spacetime possesses a coordinate expansion of order $\Delta x^{2}$ for $x$ near $x'$. Thus, it would appear that the stress-energy tensor should possess logarithmic singularities since the stress-energy tensor involves taking two derivatives of $V$. However, the coefficient of the logarithmic term in the expansion of $\hat{D}^{a}{}_{b}G_{\mathrm{S}}$ vanishes on the Cauchy horizon.

Finally, subtracting Eqs.(\ref{eq:tdiv}) from (\ref{eq:tunren}), taking the limit $\eta\to-1$ and adding the factor of $v_{1}$ according to the definition (\ref{eq:TrenDef}), we arrive at the renormalized stress-energy tensor for a massless scalar field in our quantum state on the Cauchy horizon inside the Reissner-Nordstr\"om black hole:
\begin{eqnarray}
	\label{eq:tren}
\fl	\langle \hat{T}^{\tau}{}_{\tau}\rangle&=&\frac{1}{8\pi^{2}r_{-}^{4}}\left\{B\,\xi-\frac{47 M^{2}-6 M \alpha(240\xi-193)-\alpha^{2}(2520\xi-2119)}{360 r_{-}^{2}}\right\}\nonumber\\
\fl	\langle \hat{T}^{\eta}{}_{\eta}\rangle&=&\frac{1}{8\pi^{2}r_{-}^{4}}\left\{B\,\xi+\frac{145 M^{2}+3450 M \alpha-\alpha^{2}(360\xi-6521)}{360 r_{-}^{2}}\right\}\nonumber\\
\fl	\langle \hat{T}^{\Theta}{}_{\Theta}\rangle&=&\langle \hat{T}^{\phi}{}_{\phi}\rangle=\frac{1}{16\pi^{2}r_{-}^{4}}\left\{B\, (4\xi-1)-\frac{47 M^{2}+2 M \alpha(540\xi+541)+45\alpha^{2}(56\xi+43)}{180 r_{-}^{2}}\right\}.\nonumber\\\fl
\end{eqnarray}
This is the main result. We have a closed-form representation for the stress-energy tensor on the Cauchy horizon. The components of this tensor are manifestly finite in the quantum state we have defined on the Euclidean section of the interior.

As a simple check of these results, we note that the trace for general coupling is
\begin{eqnarray}
	\fl\langle \hat{T}^{a}{}_{a}\rangle=\frac{1}{8\pi^{2}r_{-}^{4}}\left\{B\,(6\xi-1)+\frac{M^{2}-4 M \alpha(45\xi-8)+\alpha^{2}(133-720\xi)}{90 r_{-}^{2}}\right\}.
\end{eqnarray}
We can see that for conformally coupled fields $\xi=1/6$, the first term vanishes and we obtain
\begin{eqnarray}
	\langle \hat{T}^{a}{}_{a}\rangle_{\mathrm{conf}}=\frac{M^{2}+2 M\alpha+13\alpha^{2}}{720 \pi^{2}r_{-}^{6}}=\frac{v_{1}(r_{-})}{4\pi^{2}}.
\end{eqnarray}
This corresponds to the standard trace anomaly \cite{BrownOttewill1986}, as expected. A non-trivial check of our results is provided by checking that the conservation equation $\nabla_{a}\langle \hat{T}^{a}{}_{\eta}\rangle=0$ is satisfied (the other conservation equations being trivially satisfied because of the symmetries of the space-time), which is indeed the case.



\section{Conclusions and Discussion}
\label{sec:conclusions}
In this paper, we compute the regularized expectation value of the stress-energy tensor for a scalar field on the inner horizon of a Reissner-Nordstr{\"o}m black hole. Numerical calculations of the vacuum polarization on the black hole interior for the field in the Unruh and Hartle-Hawking state have been considered in Ref.~\cite{LanirPRD2019}, and they show that these states are singular on the inner horizon. If one is interested in the quantum back-reaction near the inner horizon, then it is necessary to consider the quantum field in a state that satisfies the Hadamard condition, otherwise the semi-classical approximation is violated. With this in mind, we construct the field in a quantum state that is explicitly regular on the Cauchy horizon in the sense that the Hadamard condition is satisfied for the two-point function when one of the points is on the horizon. The construction of the state involved working on a negative definite metric obtained by analytically continuing the $t$ coordinate and the polar coordinate $\theta$. Surprisingly, an exact closed-form representation of the regularized stress-energy tensor is tractable for the field in this quantum state for any value of the coupling constant.
 
There remains some interesting open questions about the calculation we present, in particular, about the quantum state we construct. For example, we have not offered any insights into what this state corresponds to physically. Is it a thermal state, for example? Presumably, this state is singular on the event horizon, though we have made no attempt to prove this. There are also some unresolved issues with the formal analysis of the double analytical continuation that we adopt and whether the two-point function has a unique continuation back to the two-point function on the Lorentzian spacetime. Certainly, there is some further work needed in these directions.

Notwithstanding the need for deeper insights into the physical interpretation of the quantum state under consideration, it is perfectly reasonable to solve the problem of computing the regularized stress-energy tensor in whatever Hadamard state is most convenient and to use the fact that differences between Hadamard states is regular to compute the stress-energy tensor in any other state. In other words, the regularization problem need only be solved in one quantum state and oftentimes the states which are most convenient to do so are those which employ Euclidean techniques. This provides a strong motivation for the approach adopted in this paper, regardless of the physical interpretation of the state.


\appendix
\section*{Appendix A: Uniform Asymptotic Series for $F(\eta)$}
\setcounter{section}{1}
In order to calculate the stress-energy tensor on the Cauchy horizon, we required a series expansion for the function we have called $F(\eta)$ (\ref{eq:FDef}) about $\eta=-1$, which in turn requires a uniform asymptotic series for the radial solution $\mathsf{p}^{1}_{\lambda}(\eta)$. In this appendix, we outline our method for achieving this; the approach is similar to a uniform asymptotic approximation developed by Candelas \cite{Candelas:1980zt} for the exterior of the Schwarzschild spacetime. The standard development of uniform asymptotics for differential equations with a large parameter is the Green-Liouville approach \cite{Olver}. This has been extended by Breen and Ottewill \cite{Breen:10} to include the radial functions on black hole spacetimes with two horizons but this method does not result in closed form representations for derivatives of the Green function near the horizon.

We start by noting that the equation satisfied by $\mathsf{p}^{1}_{\lambda}(\eta)$, in some sense, asymptotes to the equation satisfied by the conical function $\mathsf{P}^{1}_{-1/2+\rmi \lambda}(\eta)$ as $\eta\to-1$. And in particular, we have that
 \begin{equation}
	 \label{eq:pasymp}
	 \mathsf{p}^{1}_{\lambda}(\eta)\sim \frac{\pi}{\sqrt{2}\,\cosh\pi\lambda}\mathsf{P}^{1}_{-1/2+\rmi\lambda}(\eta),\qquad\eta\to-1.
\end{equation}
We look for a formal solution of the form
\begin{equation}
	\label{eq:pansatz}
\fl	\mathsf{p}^{1}_{\lambda}(\eta)=\frac{\pi}{\sqrt{2}\,\cosh\pi\lambda}\left\{\mathsf{P}^{1}_{-1/2+\rmi\,\lambda}(\eta)+g(\eta)\,\mathsf{P}_{-1/2+\rmi \lambda}(\eta)\right\}+\epsilon_{\lambda}(\eta),
\end{equation}
where $g(\eta)$ does not depend on $\lambda$. We wish to estimate the contribution of the error term $\epsilon_{\lambda}(\eta)$ in the integral that defines $F(\eta)$ near the horizon. With the particular choice
\begin{eqnarray}
	g(\eta)=\int_{-1}^{\eta}\frac{\Psi(x)}{2(1-x^{2})^{3/2}}\rmd x,\qquad \psi(\eta)=1-\left(\frac{\alpha\,\eta+M}{r_{-}}\right)^{4},
\end{eqnarray}
it can be shown that the error term satisfies
\begin{eqnarray}
	\label{eq:ErrorEqn}
	\fl \left\{\frac{\rmd}{\rmd \eta}\left(1-\eta^{2}\right)\frac{\rmd}{\rmd \eta}-\lambda^{2}-\frac{1}{4}-\frac{1}{1-\eta^{2}}\right\}\epsilon_{\lambda}(\eta)&=&-\frac{\psi(\eta)}{1-\eta^{2}}\epsilon_{\lambda}(\eta)\nonumber\\
&-&\frac{\pi}{\sqrt{2}\,\cosh\pi\lambda}\mathsf{P}_{-1/2+\rmi \,\lambda}(\eta)\,h(\eta)
\end{eqnarray}
where
\begin{equation}
\fl	h(\eta)=\left\{\frac{\rmd}{\rmd \eta}\left(1-\eta^{2}\right)\frac{\rmd}{\rmd \eta}-\frac{1-\psi(\eta)}{1-\eta^{2}}\right\}g(\eta)=\frac{4\sqrt{2} \alpha^{2}r_{+}}{5\,r_{-}^{3}}(1+\eta)^{3/2}+\Or(1+\eta)^{5/2}.
\end{equation}
Importantly, $\psi(\eta)$ has a simple zero at the Cauchy horizon so that $\psi(\eta)/(1-\eta^{2})$ is regular there. Moreover, $h(\eta)$ is bounded on a neighbourhood of this point.  This equation can be solved formally by the method of variation of parameters
\begin{equation}
	\label{eq:ErrorInt}
	\epsilon_{\lambda}(\eta)=\int_{0}^{\eta}K_{\lambda}(\eta, x)\left(\tilde{\psi}(x)\,\epsilon_{\lambda}(x)+\tilde{\mathsf{P}}_{\lambda}(x)\,h(x)\right)\,dx,
\end{equation}
where the kernel $K_{\lambda}(\eta, x)$ is defined by
\begin{equation}
	\label{eq:KernelDef}
\fl	K_{\lambda}(\eta, x)=\frac{|\Gamma(-\frac{1}{2}+\rmi\,\lambda)|^{2}}{2}\left\{\mathsf{P}_{-1/2+\rmi\,\lambda}^{1}(\eta)\,\mathsf{P}_{-1/2+\rmi\,\lambda}^{1}(-x)-\mathsf{P}_{-1/2+\rmi\,\lambda}^{1}(x)\,\mathsf{P}_{-1/2+\rmi\,\lambda}^{1}(-\eta)\right\},
\end{equation}
and we have simplified the notation by identifying
\begin{equation}
	\tilde{\psi}(\eta)=\frac{\psi(\eta)}{1-\eta^{2}},\qquad \tilde{\mathsf{P}}_{\lambda}(\eta)=\frac{\pi}{\sqrt{2}\,\cosh\pi\lambda}\mathsf{P}_{-1/2+\rmi\,\lambda}(x).
\end{equation}
It is straightforward to show that (\ref{eq:ErrorInt}) is a solution to Eq.~(\ref{eq:ErrorEqn}) using the Wronskian
\begin{equation}
	(1-\eta^{2})\,W\left\{\mathsf{P}^{1}_{-1/2+\rmi\lambda}(\eta),\mathsf{P}^{1}_{-1/2+\rmi\lambda}(-\eta)\right\}=\frac{2}{\Gamma(-\frac{1}{2}+\rmi\lambda)\,\Gamma(-\frac{1}{2}-\rmi\lambda)}.
\end{equation}
Now uniqueness and boundedness of the solution (\ref{eq:ErrorInt}) is guaranteed for general integral equations of the type (\ref{eq:ErrorEqn}) (see Theorem 10.1, Chapter 6 of Ref.~\cite{Olver}) provided the following assumptions hold:
\begin{enumerate}
	\item The functions $\tilde{\mathsf{P}}_{\lambda}(x)$, $h(x)$ and $\tilde{\psi}(x)$ are continuous on $x\in (-1,\beta)$ save for a finite number of discontinuities or infinities.
	\item The kernel $K_{\lambda}(\eta,x)$ and its first two partial $\eta$ derivatives are continuous functions of both $x$ and $\eta$ on $(-1,\beta)$.
	\item $K_{\lambda}(\eta,\eta)=0$.
	\item For $\eta\in(-1,\beta)$ and $x\in (-1,\eta]$, there exists positive continuous functions $\mathcal{P}^{(j)}_{\lambda}(\eta)$ and a continuous function $\mathcal{Q}_{\lambda}(x)$ such that
	\begin{equation*}
		\fl|K_{\lambda}(\eta,x)|\le \mathcal{P}^{(0)}_{\lambda}(\eta)\mathcal{Q}_{\lambda}(x),\quad \left|\frac{\partial K_{\lambda}(\eta,x)}{\partial \eta}\right|\le \mathcal{P}^{(1)}_{\lambda}(\eta)\mathcal{Q}_{\lambda}(x),\quad \left|\frac{\partial^{2}K_{\lambda}(\eta,x)}{\partial\eta^{2}}\right|\le \mathcal{P}^{(2)}_{\lambda}(\eta)\mathcal{Q}_{\lambda}(x).
	\end{equation*}
	\item When $\eta\in (-1,\beta)$, the following integrals converge
	\begin{equation}
		\Phi(\eta)=\int_{-1}^{\eta}|h(x)|dx,\quad \Psi(\eta)=\int_{-1}^{\eta}|\tilde{\psi}(x)|dx,
	\end{equation}
	and the following suprema are finite
	\begin{equation}
		\delta\equiv \sup\{\mathcal{Q}_{\lambda}(\eta)|\tilde{\mathsf{P}}_{\lambda}(\eta)|\},\quad \delta_{0}\equiv \sup\{\mathcal{P}^{(0)}_{\lambda}(\eta)\,\mathcal{Q}_{\lambda}(\eta)\}.
	\end{equation}
\end{enumerate} 
These conditions do indeed hold in our case for $-1<\beta<1$ though it remains to find explicit $\mathcal{P}^{(j)}_{\lambda}(\eta)$ and $\mathcal{Q}_{\lambda}(\eta)$ satisfying condition (iv). Let us derive explicitly only the first bound in (iv), the others following a similar route. We start by noting that, for fixed $\eta$, the product $\mathsf{P}^{1}_{-1/2+\rmi \lambda}(x)\mathsf{P}^{1}_{-1/2+\rmi \lambda}(-\eta)$ is a monotonically decreasing function of $x$ tending to $\infty$ as $x\to-1$, while $\mathsf{P}^{1}_{-1/2+\rmi \lambda}(\eta)\mathsf{P}^{1}_{-1/2+\rmi \lambda}(-x)$ is monotonically increasing over $(0,\infty)$ as $x$ ranges over $(-1,\eta)$. This implies that
\begin{equation*}
	\mathsf{P}^{1}_{-1/2+\rmi \lambda}(\eta)\mathsf{P}^{1}_{-1/2+\rmi \lambda}(-x)-\mathsf{P}^{1}_{-1/2+\rmi \lambda}(x)\mathsf{P}^{1}_{-1/2+\rmi \lambda}(-\eta)>0,\quad x<\eta.
\end{equation*}
This trivially implies a positive kernel $K(\eta, x)>0$ for $x<\eta$. Moreover, since each product in this difference is positive, we have
\begin{equation}
	0<K(\eta, x)<\frac{|\Gamma(-\frac{1}{2}+\rmi\,\lambda)|^{2}}{2}\mathsf{P}^{1}_{-1/2+\rmi \lambda}(\eta)\mathsf{P}^{1}_{-1/2+\rmi \lambda}(-x),\qquad x<\eta.
\end{equation}
Hence, the first inequality in condition (iv) above is satisfied with
\begin{equation}
\fl	\mathcal{P}^{(0)}_{\lambda}(\eta)=\frac{\pi}{\sqrt{2}\,\cosh\pi\lambda}\mathsf{P}_{-1/2+\rmi\lambda}^{1}(\eta)=\tilde{\mathsf{P}}_{\lambda}(\eta),\quad \mathcal{Q}_{\lambda}(x)=\frac{1}{\sqrt{2}\,(\lambda^{2}+1/4)}\mathsf{P}^{1}_{-1/2+\rmi \lambda}(-x),
\end{equation}
where we have used the fact that $|\Gamma(-1/2+\rmi\,\lambda)|^{2}=\pi\, \sech(\pi\lambda)/(\lambda^{2}+1/4)$. Bounding derivatives of the kernel is identical except the $\mathcal{P}_{\lambda}^{(j)}(\eta)$ ($j=1,2$) involve derivatives of the conical functions.

With these particular choices, it is now also a straightforward matter to explicitly compute the suprema $\delta$ and $\delta_{0}$. In particular, using the monotonicity of the conical functions and the asymptotic forms
\begin{eqnarray}
	\mathsf{P}^{1}_{-1/2+\rmi\,\lambda}(\eta)\sim\frac{\sqrt{2}}{\pi}\cosh \pi\lambda\,(1+\eta)^{-1/2},\qquad\eta\to-1,\nonumber\\
	\mathsf{P}^{1}_{-1/2+\rmi\,\lambda}(-\eta)\sim\frac{1}{\sqrt{2}}(\lambda^{2}+1/4)(1+\eta)^{1/2},\qquad \eta\to-1,
\end{eqnarray}
we obtain $\delta=\delta_{0}=1/2$.

Finally, the theorem which guarantees uniqueness and boundedness of the error also gives the explicit bound
\begin{equation}
	\frac{|\epsilon_{\lambda}(\eta)|}{\mathcal{P}^{(0)}_{\lambda}(\eta)}\le \delta \Phi (\eta)\exp\{\delta_{0}\Psi(\eta)\}.
\end{equation}
Hence, we get the following uniform estimate for the contribution of the error near the Cauchy horizon,
\begin{equation}
	\epsilon_{\lambda}(\eta)\sim \tilde{\mathsf{P}}_{\lambda}(\eta)(1+\eta)^{5/2},\quad \eta\to-1,
\end{equation}
using the fact that $\Phi(\eta)\sim (1+\eta)^{5/2}$ for $\eta\to -1$. From this we can estimate the contribution of this error to the function $F(\eta)$ defined by (\ref{eq:FDef}), that is,
\begin{eqnarray}
	\fl\int_{0}^{\infty}2\lambda\,\tanh\pi\lambda\,\epsilon_{\lambda}(\eta)d\lambda&\sim& (1+\eta)^{5/2}\int_{0}^{\infty}\frac{\pi \lambda\, \tanh\pi\lambda}{\cosh\pi\lambda}\mathsf{P}_{-1/2+\rmi\,\lambda}^{1}(\eta)\,\rmd \lambda \quad\eta\to-1\nonumber\\
\fl	&=&(1+\eta)(1-\eta)^{1/2},
\end{eqnarray}
where the last line follows by differentiating the identity (\ref{eq:Heine}). Hence the error term does not contribute to $(1+\eta)^{-1/2}F(\eta)$ in the limit as the Cauchy horizon is approached. Only the first two terms in (\ref{eq:pansatz}) contribute and are easily calculated again from the identity (\ref{eq:Heine}). The result is
\begin{equation}
\fl	F(\eta)=\frac{2}{(1+\eta)^{3/2}}-\frac{(M+3\alpha)}{2\,r_{-}\,(1+\eta)^{1/2}}-\frac{(M+3\alpha)^{2}}{16 r_{-}^{2}}(1+\eta)^{1/2}+\Or(1+\eta).
\end{equation}

As a final note in this appendix, we point out that (\ref{eq:pansatz}) is not the most general asymptotic form for $\mathsf{p}^{1}_{\lambda}(\eta)$ and in particular, there is a freedom to add multiples of the subdominant solution $\beta_{\lambda}\,\mathsf{q}^{1}_{\lambda}(\eta)$. The $\beta_{\lambda}$ coefficients are chosen in such a way that $\mathsf{p}^{1}_{\lambda}(\eta)$ satisfies the desired boundary conditions at the event horizon $\eta=1$. In the next appendix, we outline how to calculate the $\beta_{\lambda}$ which correspond to $p^{1}_{\lambda}(\eta)\to 0$ at the event horizon. In any case, for unspecified boundary conditions, we have
\begin{equation}
\fl	F(\eta)=\frac{2}{(1+\eta)^{3/2}}-\frac{(M+3\alpha)}{2\,r_{-}\,(1+\eta)^{1/2}}-\left(\frac{(M+3\alpha)^{2}}{16 r_{-}^{2}}-B\right)(1+\eta)^{1/2}+\Or(1+\eta),
\end{equation}
where
\begin{equation}
	\label{eq:BDef}
	B=\int_{0}^{\infty}2\lambda\,\tanh\pi\lambda\,\beta_{\lambda}\,\rmd\lambda.
\end{equation}
For the state to be regular, we require that the constant $B$ be finite, that is, $\beta_{\lambda}\sim \rm{o}(\lambda^{-2})$ for large $\lambda$.

\section*{Appendix B: Evaluating the $\beta_{\lambda}$ Coefficients}
\setcounter{section}{2}
We describe how to evaluate the $\beta_{\lambda}$ coefficients appearing in the definition of the constant $B$ for the case where we impose the boundary condition $\mathsf{p}^{1}_{\lambda}(\eta)\to 0$ at the event horizon.

We begin with the Wronskian relation between the $n=\pm1$ radial functions of the first and second kind,
\begin{equation}
\mathsf{q}^{1}_{\lambda}(\eta)\frac{\rmd}{\rmd\eta}\mathsf{p}^{1}_{\lambda}(\eta)-\mathsf{p}^{1}_{\lambda}(\eta)\frac{\rmd}{\rmd\eta}\mathsf{q}^{1}_{\lambda}(\eta)=-\frac{2}{1-\eta^{2}}.
\end{equation}
Dividing across by $(\mathsf{q}^{1}_{\lambda})^{2}$ and integrating we obtain the following integral expression for $\mathsf{p}^{1}_{\lambda}(\eta)$
\begin{equation}
\label{eq:pint}
\mathsf{p}^{1}_{\lambda}(\eta)=2\,\mathsf{q}^{1}_{\lambda}(\eta)\int_{\eta}^{1}\frac{\rmd x}{(1-x^{2})[\mathsf{q}^{1}_{\lambda}(x)]^{2}}\,\, ,
\end{equation}
where we have used the boundary condition $\mathsf{p}^{1}_{\lambda}(\eta)\rightarrow0$ as $\eta\rightarrow 1$ to fix the upper bound of the integral. A standard Frobenius series expansion for $\mathsf{q}^{1}_{\lambda}(x)$ about $x=-1$ yields
\begin{equation}
\label{eq:qseries}
\mathsf{q}^{1}_{\lambda}(x)=(x+1)^{1/2}+\left(\frac{5 M+3\alpha}{16\,r_{-}}+\frac{\lambda^{2}}{4}\right)(x+1)^{3/2}+\Or(x+1)^{5/2}
\end{equation}
and therefore we have
\begin{equation}
\label{eq:qinvseries}
\frac{1}{[\mathsf{q}^{1}_{\lambda}(x)]^{2}}=\frac{1}{x+1}-\left(\frac{5 M+3\alpha}{8\,r_{-}}+\frac{\lambda^{2}}{2}\right)+\Or(x+1).
\end{equation}
We wish to subtract and add this singular behaviour from the integrand (\ref{eq:pint}) so that we isolate the divergences at $\eta=-1$. However, we do not subtract the terms on the right-hand side of (\ref{eq:qinvseries}) over the entire integration range since (\ref{eq:qinvseries}) has a non-integrable singularity at the upper bound $\eta=1$. Instead we write (\ref{eq:pint}) as
\begin{eqnarray}
\fl \mathsf{p}^{1}_{\lambda}(\eta)&=&\mathsf{q}^{1}_{\lambda}(\eta)\int_{\eta}^{0}\frac{2}{(1-x^{2})}\left(\frac{1}{[\mathsf{q}^{1}_{\lambda}(x)]^{2}}-\frac{1}{(x+1)}+\left(\frac{5 M+3\alpha}{8\,r_{-}}+\frac{\lambda^{2}}{2}\right)\right)\rmd x \nonumber\\
\fl&&+\mathsf{q}^{1}_{\lambda}(\eta)\int_{\eta}^{0}\frac{2}{(1-x^{2})}\left(\frac{1}{(x+1)}-\left(\frac{5 M+3\alpha}{8\,r_{-}}+\frac{\lambda^{2}}{2}\right)\right)\rmd x\nonumber\\
\fl&&+\mathsf{q}^{1}_{\lambda}(\eta)\int_{0}^{1}\frac{2}{(1-x^{2})[\mathsf{q}^{1}_{\lambda}(x)]^{2}}\rmd x.
\end{eqnarray}
The first integral on the righthand side above converges and is amenable to a Taylor series about $\eta=-1$ while the second term can be integrated explicitly. Using (\ref{eq:qseries}), we obtain
\begin{eqnarray}
\label{eq:p1series1} \fl \mathsf{p}^{1}_{\lambda}(\eta)&=&\frac{1}{(\eta+1)^{1/2}}+\left(\frac{M+7\alpha}{8\,r_{-}}+\frac{\lambda^{2}}{2}\right)(\eta+1)^{1/2}\ln\left(\frac{\eta+1}{2}\right) \nonumber\\
 \fl &+&\left\{ I_{\lambda}+J_{\lambda}-(1-\case{1}{4}\lambda^{2})+\frac{5M+3\alpha}{16\,r_{-}}\right\}(\eta+1)^{1/2}+\Or((\eta+1)^{3/2}\ln(\eta+1)),
\end{eqnarray}
where
\begin{eqnarray} I_{\lambda}&=&\int_{-1}^{0}\frac{2}{(1-x^{2})}\left(\frac{1}{[\mathsf{q}^{1}_{\lambda}(x)]^{2}}-\frac{1}{(x+1)}+\left(\frac{5 M+3\alpha}{8\,r_{-}}+\frac{\lambda^{2}}{2}\right)\right)\rmd x,\nonumber\\
J_{\lambda}&=&\int_{0}^{1}\frac{2}{(1-x^{2})[\mathsf{q}^{1}_{\lambda}(x)]^{2}}\rmd x.
\end{eqnarray}

Recall that we also have an alternate expression for $\mathsf{p}^{1}_{\lambda}(\eta)$ which effectively defines the $\beta_{\lambda}$ coefficients we are trying to compute,
\begin{equation}
	\fl	\mathsf{p}^{1}_{\lambda}(\eta)=\frac{\pi}{\sqrt{2}\,\cosh\pi\lambda}\left\{\mathsf{P}^{1}_{-1/2+\rmi\,\lambda}(\eta)+g(\eta)\,\mathsf{P}_{-1/2+\rmi \lambda}(\eta)\right\}+\epsilon_{\lambda}(\eta)+\beta_{\lambda}\,\mathsf{q}^{1}_{\lambda}(\eta).
\end{equation}
Each term here is straightforward to expand about $\eta=-1$ resulting in an equivalent series for $\mathsf{p}^{1}_{\lambda}(\eta)$,
\begin{eqnarray}
\label{eq:p1series2}
\fl\mathsf{p}^{1}_{\lambda}(\eta)&=&\frac{1}{(\eta+1)^{1/2}}+\left(\frac{M+7\alpha}{8\,r_{-}}+\frac{\lambda^{2}}{2}\right)(\eta+1)^{1/2}\ln\left(\frac{\eta+1}{2}\right) \nonumber\\
\fl &+&\left( \beta_{\lambda}+\frac{1-4\lambda^{2}}{8}+\left(\frac{M+7\alpha}{8 \,r_{-}}+\frac{\lambda^{2}}{2}\right)(H_{-1/2+\rmi\,\lambda}+H_{-1/2-\rmi\,\lambda})\right)(\eta+1)^{1/2}\nonumber\\
\fl&+&\Or((\eta+1)^{3/2}\ln(\eta+1))\,,
\end{eqnarray}
where $H_{z}$ is the Harmonic number.

Comparing our two equivalent series expansions (\ref{eq:p1series1}) and (\ref{eq:p1series2}) yields an expression for $\beta_{\lambda}$ that is numerically tractable:
\begin{equation}
\fl\beta_{\lambda}= I_{\lambda}+J_{\lambda}-\frac{3}{4}\left(\frac{3}{2}-\lambda^{2}\right)+\frac{5 M+3\alpha}{16\,r_{-}}-\left(\frac{M+7\alpha}{8\,r_{-}}+\frac{\lambda^{2}}{2}\right)(H_{-1/2+\rmi\,\lambda}+H_{-1/2-\rmi\,\lambda}).
\end{equation}

The integral $I_{\lambda}$ is most effectively calculated by employing a high-order series solution to $\mathsf{q}^{1}_{\lambda}(\eta)$ in order to cancel the divergences explicitly in the integrand at the lower bound of the integral. The integral can then be performed accurately and efficiently. Computing the integral $J_{\lambda}$ requires the full numerical solution for $\mathsf{q}^{1}_{\lambda}(\eta)$. Nevertheless, given the numerically computed modes, the integral is straightforward to compute numerically modulo some numerical instability very close to the upper bound of the integral. However, this does not present a problem since the integral cuts off exponentially as the upper bound is approached. In practice, we cut off the numerical integral at $x=1-\epsilon$ with $\epsilon=10^{-6}$ with little loss of accuracy.

We find numerically that $\beta_{\lambda}\sim \lambda^{-4}$ for large $\lambda$ and so the integral (\ref{eq:BDef}) converges quickly, though the speed of convergence is sensitive to the black hole parameters. In particular, for fixed $M$, the convergence is slower for increasing $\alpha=\sqrt{M^{2}-Q^{2}}$. We compute $\beta_{\lambda}$ for $\lambda\in[0,5]$ with a mesh size of $0.1$, $\lambda \in [5,40]$ with a mesh size of unity and $\lambda\in[40,100]$ with a mesh size of 5. We then use Mathematica's inbuilt Interpolation routine to generate an interpolating function for $\beta_{\lambda}$. Finally we numerically integrate this interpolating function (cutting off the integral at $\lambda=100$) to obtain $B$ to  3-5 decimal places of accuracy. The table above shows our computed values for $M=1$ and a range of $\alpha$ values, rounded to 4 decimal places. One can fit these computed values to plot $\beta_{\lambda}$ as a function of $\alpha$ for all $\alpha$ values. This is plotted below. Note that $\beta_{\lambda}$ becomes negative only near extremality $\alpha=0$.

\begin{table}
 \caption{The integral $B$ corresponding to boundary condition $\mathsf{p}^{1}_{\lambda}(\eta)\to 0$ as $\eta\to 1$, with black hole parameters $M=1$ and various $\alpha=\sqrt{M^{2}-Q^{2}}$ values.}
\begin{center}
\begin{tabular}{| c||c | c | c | c | c | c|} \hline
   $\alpha$& 1/10   &    1/5   &    3/10     &     2/5    &     1/2    &    3/5   \\
\hline
$B$ &0.0723                        & 0.2107 &     0.4684                     &     0.9524                          &     1.9002                                 &          3.9137    \\
\hline
\end{tabular}
\end{center}
\label{default}
\end{table}

\begin{figure}[h!]
	\begin{center}
	\includegraphics[width=11cm]{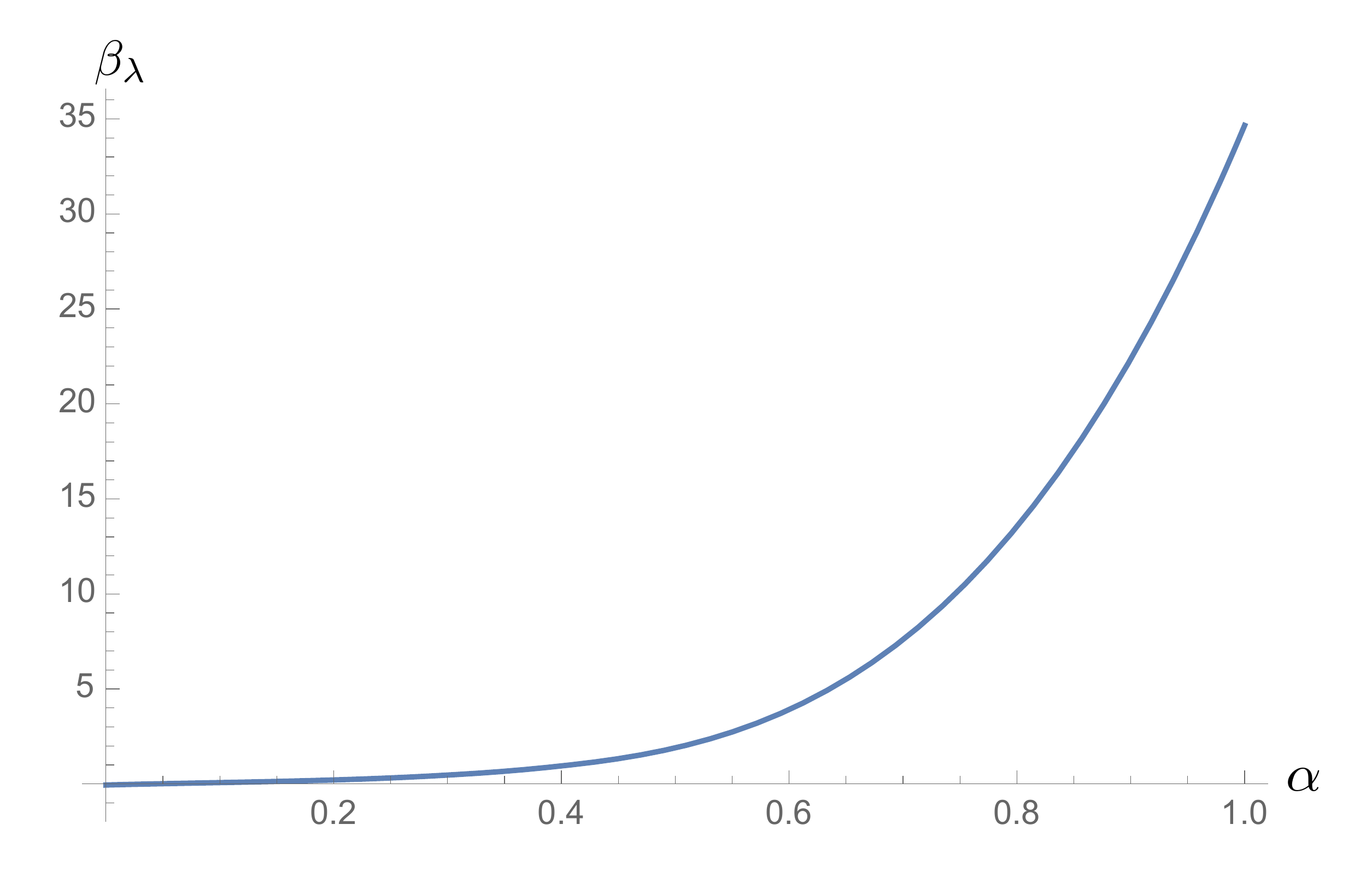}
	\caption{Plot of $\beta_{\lambda}$ with $M=1$ as a function of $\alpha=\sqrt{1-Q^{2}}$.}
	\end{center}
\end{figure}
\section*{Acknowledgements}
I thank Amos Ori and Cormac Breen for sharing their insights and for helpful suggestions.

\section*{References}
\bibliographystyle{iopart-num}
\bibliography{my_bib}

\newpage

\end{document}